\documentclass[preprint]{aastex}

\begin{document}

\title{The IBIS soft gamma-ray sky after 1000 INTEGRAL orbits\footnote{\rm Based on observations with INTEGRAL, an ESA project
with instruments and science data centre funded by ESA member states
(especially the PI countries: Denmark, France, Germany, Italy,
Switzerland, Spain), Czech Republic and Poland, and with the
participation of Russia and the USA.}}

\author{A. J. Bird\altaffilmark{a},
    A. Bazzano\altaffilmark{b},
    A. Malizia\altaffilmark{c},   
    M. Fiocchi\altaffilmark{b},
    V. Sguera\altaffilmark{c},
    L. Bassani\altaffilmark{c},
    A. B. Hill\altaffilmark{a},
    P. Ubertini\altaffilmark{b},
    C. Winkler\altaffilmark{d}
    }

\altaffiltext{a}{School of Physics and Astronomy, University of Southampton, SO17 1BJ, UK}
\altaffiltext{b}{IAPS/INAF, Italy}
\altaffiltext{c}{IASF/INAF, Bologna, Italy}
\altaffiltext{d}{ESA-ESTEC, Research and Scientific Support Dept., Keplerlaan 1, 2201 AZ, Noordwijk, The Netherlands}


\begin{abstract}

We report here an all-sky soft gamma-ray source catalog based on IBIS observations performed during the first 1000 orbits of INTEGRAL. The database for the construction of the source list consists of all good quality data available from launch in 2002 up to the end of 2010. This corresponds to $\sim$110 Ms of scientific public observations with a concentrated coverage on the Galactic Plane and extragalactic deep exposures. This new catalog includes 939 sources above a 4.5 sigma significance threshold detected in the 17-100 keV energy band, of which 120 represent previously undiscovered soft gamma-ray emitters. The source positions are determined, mean fluxes are provided in two main energy bands, and are reported together with the overall source exposure. Indicative levels of variability are provided, and outburst times and durations are given for transient sources. Comparison is made with previous IBIS catalogs, and those from other similar missions. 
\end{abstract}



\section{Introduction}

More than eleven  years of observations in the energy range from 5 keV up to 10 MeV have been performed with the INTEGRAL observatory, which was selected as the M2 mission within ESA's Horizon 2000 programme. The  observing time of INTEGRAL is awarded competitively via a general programme which is open  to the community at large, and includes Targets of Opportunity, normal observations and Key Programmes. The latter category consists of deep observations requiring a few Ms observing time, and may accommodate various different requests from the observer community for amalgamated single or multiple targets within the selected sky fields. Typical observation times range between 100~ks and more than two weeks, and a number of the programmes have provided regular monitoring of the Galaxy by returning to the same area of sky on multiple occasions.

Survey observations with INTEGRAL make full use of the large field of view of the IBIS coded mask telescope, one of the two main instruments on board. IBIS, with its large field of view (28 x 28$^\circ$, 9 x 9$^\circ$ fully coded), excellent imaging and spectral capability is ideal for survey work \citep{Ubertini2003}. The imaging system provides a location accuracy of 0.5-4$'$ depending on the source strength. For the large numbers of newly-detected unidentified sources, these localizations are sufficiently good  to enable searches for their soft X-ray counterparts. Results presented here are derived from ISGRI (Lebrun et al., 2003), the low energy array on IBIS, a pixelated CdTe detector operating in the energy band 17-1000 keV. 

Since 2004, a sequence of IBIS survey catalogs \citep{Bird2004cat1,Bird2006cat2,Bird2007cat3,Bird2010cat4} based on data from the ISGRI detector system have been published at regular intervals,  making use of an ever-increasing dataset as new observations become publicly available. The last edition of the IBIS survey \citep{Bird2010cat4}, comprising 723 sources, was released in 2010, and was based on INTEGRAL data collected between 2003 February and 2008 April.  
The overall content of this unbiased catalog comprised known AGNs (35\% ), X-ray binaries (31\%), pulsars and other sources (5\%) while 29\% of the sources were unknown or detected for the first time with INTEGRAL. A large number of observations at X-ray wavelengths with {\em Swift},  XMM and Chandra followed in order to obtain better position determinations and hence a more reliable optical identification.  
Other INTEGRAL-based catalogs have been produced, and focus on specific sky areas such as the Galactic Plane \citep{Krivonos2012, Krivonos2010} or on specific source classes \citep{Revnivtsev2008a, Lutovinov2007, Lutovinov2013, Bassani2006, Malizia2012, Revnivtsev2008b, Scaringi2010a, Beckman2009, Sazonov2007}.  Another catalog produced in 2008 \citep{Bouchet2008} was based on SPI (the other primary wide-field instrument on INTEGRAL) observations. Apart from just one variable source, all the objects listed in that publication were included in the 4th IBIS/ISGRI survey catalog (hereafter ``cat4") \citep{Bird2010cat4}.

The most recent INTEGRAL survey (Krivonos et al, 2012) is based on nine years of averaged sky images and lists only those sources detected along the Galactic Plane ($|b| <17.5^\circ$)  in three energy bands (17-60, 17-35 and 35-80 keV); it includes 402 objects exceeding a 4.7$\sigma$ detection threshold on the nine years average map.

In all, the total number of INTEGRAL-discovered sources (i.e. those with an IGR designation) from the various catalogs up to the end of 2013 consists of $\sim$560 IGR detections, of which only 39\% remain unidentified.  In large part this unidentified fraction can be attributed to transient sources for which rapid follow-up was not available.

A noteworthy innovation is the SIX catalog \citep{Bottacini2012} based on a new approach developed to survey the sky at hard X-ray energies (18-55 keV energy band) by combining the observations of {\it Swift}/BAT and INTEGRAL/IBIS to enhance the exposure time and reduce systematic uncertainties. This survey may be considered a survey from a {\em virtual} new hard X-ray mission, and should provide higher sensitivity than individual instrument surveys.  The method  has been applied to 6200 deg$^2$ of extragalactic sky ($\sim$20\% of the entire extragalactic sky) and lists 113 sources mostly of extragalactic nature: 91 AGNs, 2 clusters of galaxies, 3 Galactic sources, 3 previously detected  X-ray sources, and 14 unidentified sources. Suppression of systematics is a key feature of this method, and no false detections due to statistical or systematic fluctuations are expected by the authors. 

Here we present an update to the 4th IBIS/ISGRI catalog with data collected up to INTEGRAL orbit 1000, i.e. up to the end of 2010, that now comprises over 900 sources. For this updated database, we again made use  of the `bursticity' tool to improve detection of sources showing high variability and provide enhanced weak transient source detection. In particular, we use improved algorithms to provide a critical re-analysis of the methods used in Bird et al.(2010) and give additional quality flagging in order to reduce the expected levels of false detections in this new work. Details on the analysis and production of this new catalog (hereafter ``this work" or ``cat1000") are in sections \ref{sec:analysis} and \ref{sec:tabledesc}, and comparison with the more recent similar catalogs is provided in section \ref{sec:results}.


\section{Data analysis and catalog construction}
\label{sec:analysis}
 
\subsection{Input dataset}

For this work, all publicly available INTEGRAL data obtained up to the end of 2010 has been processed. This may be compared to cat4,  that used public data up to April 2007 (plus Public and Core Programme data up to April 2008).  During each satellite orbit (revolution; approximately three days) INTEGRAL operates by dividing each observation into a sequence of short pointings (science windows or scw) with a typical duration of 2 ks. Our dataset extends from revolution twelve onwards, and includes the performance verification, calibration, original core programme (including Galactic Centre Deep Exposure and Galactic Plane Scans)  and all pointed observations selected in the various observer AO phases up to revolution 1000 (December 2010). All data from revolution twelve onwards were processed unless flagged as Bad Time Intervals (flagging is provided by the INTEGRAL Science Data Centre, ISDC) for a total of $\sim$73000 scws (cf. 39548 for cat4). The input catalog used was the INTEGRAL reference catalog v31, that includes all sources of the 4th IBIS/ISGRI catalog, further updated for any new INTEGRAL detected sources published via papers and ATELs since 2010. A final cleaning catalog containing all previously declared INTEGRAL-detected sources was created and used as the input catalog for all pipeline processing.

\begin{figure}[htbp]
\centering
\includegraphics[width=0.75\columnwidth]{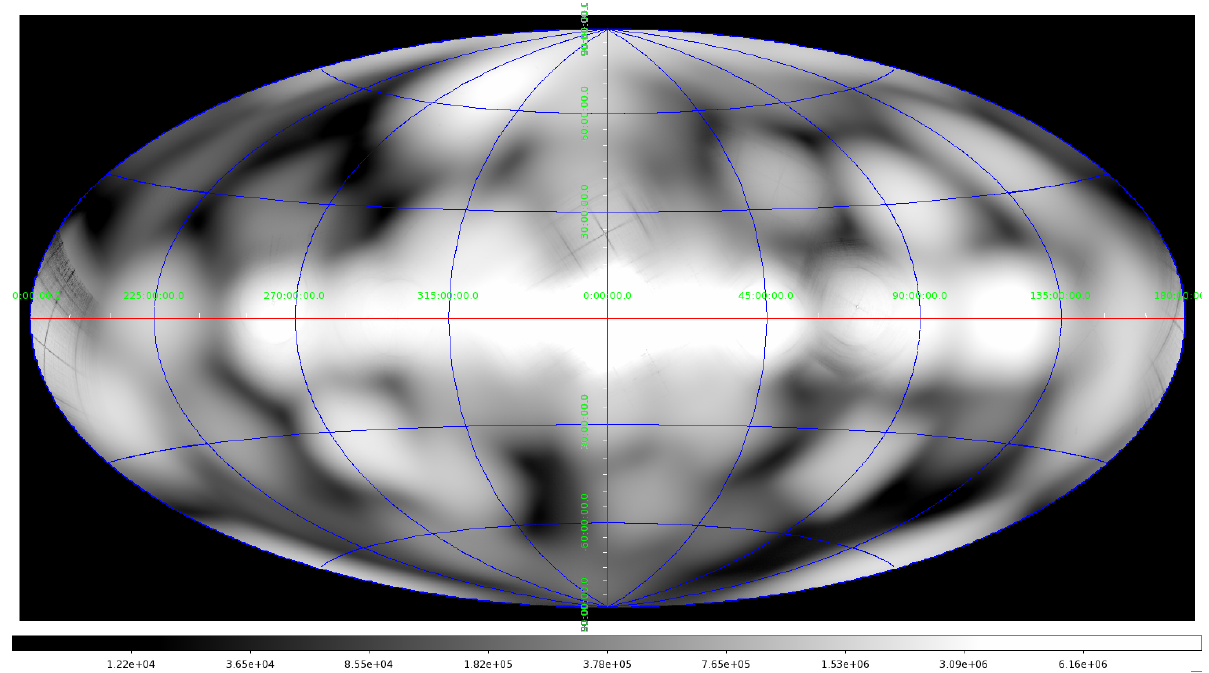}
\vspace{0.5cm}
\includegraphics[width=0.75\columnwidth]{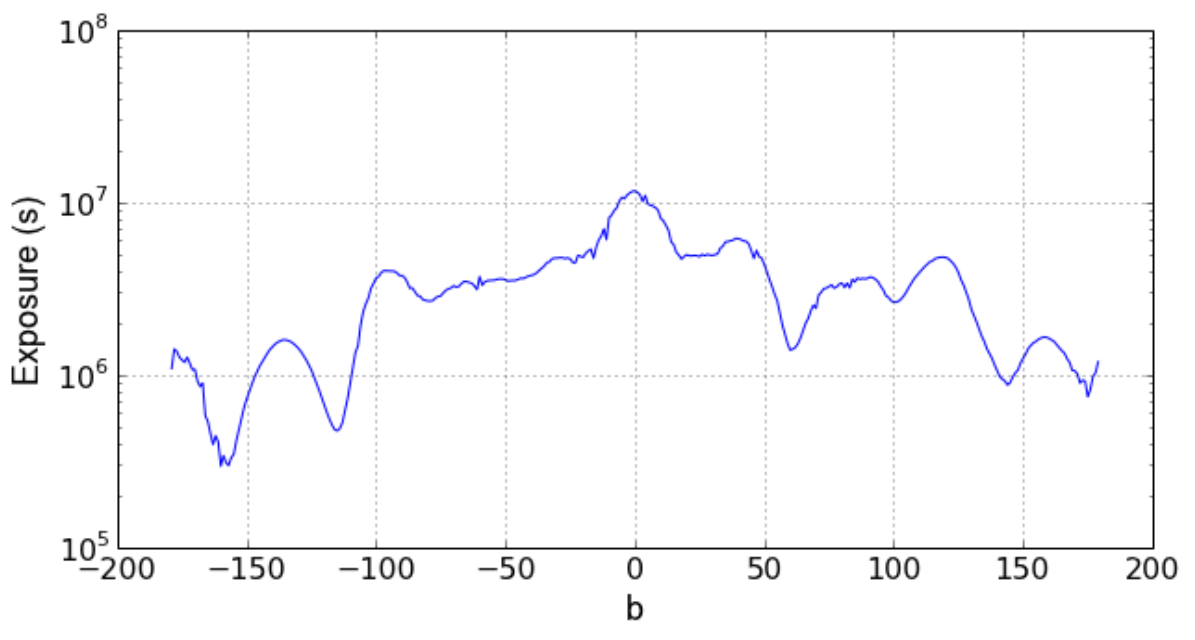}
\caption{(upper) Overall exposure map for 18-60 keV processing; the features of residual filtering (section~\ref{sec:mosaics}) can be seen around the positions of the brightest sources. (lower) The exposure along the Galactic Plane resulting from all exposures, but dominated by the Galactic Bulge and Galactic Plane Scan programmes.
\label{fig:expo}}
\end{figure}

The total exposure in the dataset (the sum of exposures of all initially selected scw) is 124 Ms (cf. 70 Ms in cat4), and the resulting all-sky exposure distribution is shown in Figure~\ref{fig:expo} (upper).

As a result of the Core Programme and Key Programmes that operated from 2006, the Galactic Bulge is well covered with $\sim$12 Ms and the entire Galactic Plane has a coverage of at least 300~ks, rising dramatically in areas where specific sources have been targeted. The Galactic anti-centre region has been less well covered due to mission planning constraints as this region is competing with the Galactic Centre for observing time. The exposure profile along the Galactic Plane is shown in Figure~\ref{fig:expo} (lower). 

\begin{figure}[htbp]
\centering
\includegraphics[width=0.75\columnwidth]{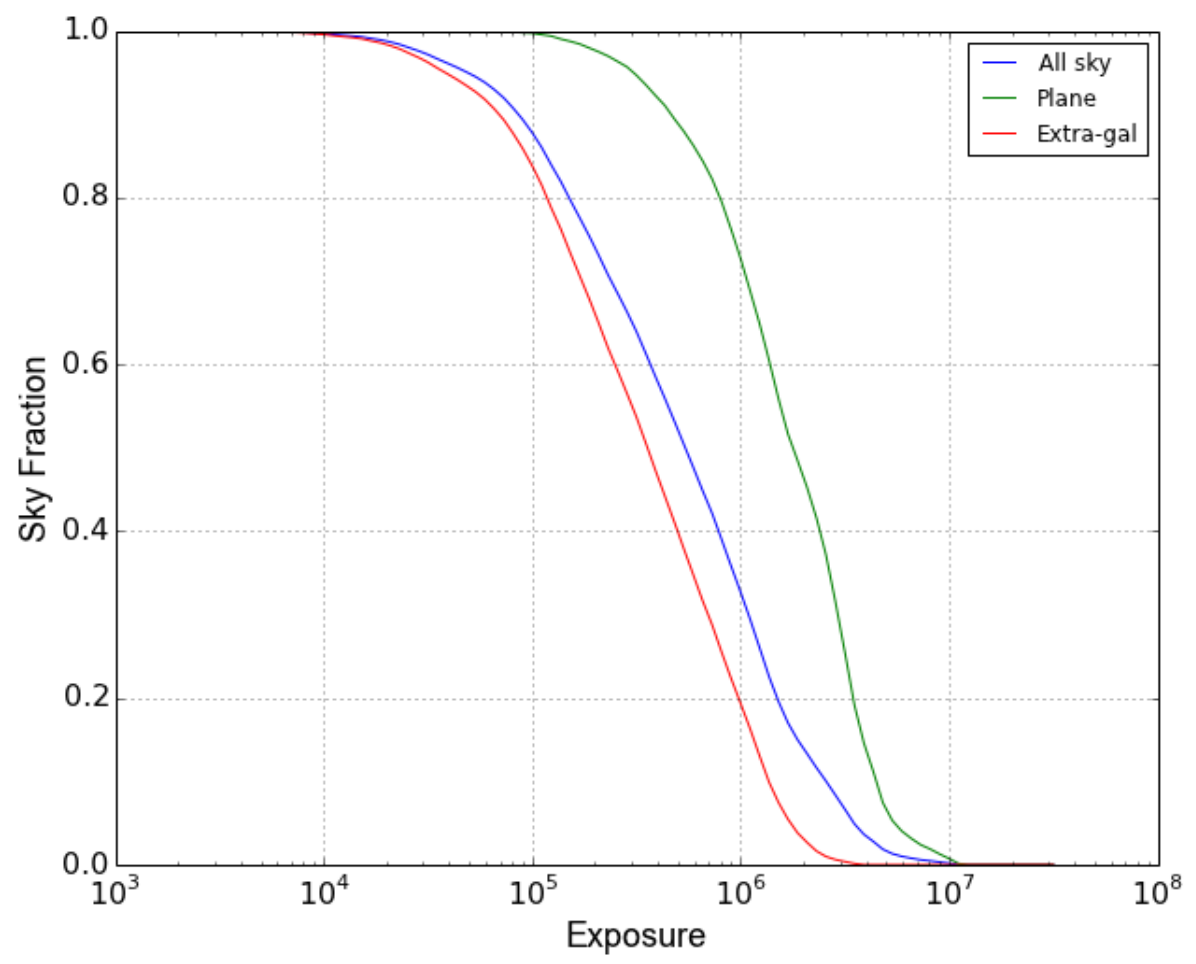}
\vspace{0.5cm}
\caption{Fraction of sky covered as a function of exposure time.
\label{fig:expo_frac}}
\end{figure}

The fraction of sky exposed to a certain level is shown in Figure~\ref{fig:expo_frac} which emphasises the different exposure patterns in the Galactic Plane ($|b|<15^{\circ}$) and the extra-galactic sky($|b|>15^{\circ}$). In the Galactic Plane, 75\% of the sky is covered to better than 1 Ms, while only 20\% of the extragalactic sky is covered to the same level. Overall, around one third of the sky is covered to 1 Ms level, and 90\% of the sky is covered to 100 ks.

\subsection{Data processing and pipeline processing \label{pipeline}}
  
The data processing was performed with the Standard {\it {OSA 9.0}} software to produce sky images of individual scws in five different energy bands (17-30, 30-60, 18-60, 20-40 and 20-100 keV). For mosaic construction, source searching, candidate list production, and final source selection we have been largely following the 4th IBIS catalog procedure as described in Bird et al, 2010. In the following sections, we only report the main changes with respect to the earlier methodology.

\subsubsection{Mosaic construction} 

\label{sec:mosaics}
Each scw image was tagged with its rms (after removal of sources) to act as an indicator of overall image quality. The primary aim of this step is to remove data taken during periods of enhanced background during solar activity or soon after perigee passage. Filtering was applied based on the rms value of the image background, such that the rms should not exceed a limit of 2$\sigma$ above the mean image rms for the whole dataset. This function  somewhat overlaps with the BTI flagging provided by the ISDC, but we still removed around 5400 of the science windows that exceed this rms limit, for a loss of 11 Ms or 8\% to the total exposure.  We note that there is a clearly increasing trend in the image rms throughout the mission, especially in the energy bands extending below 20 keV. In future analyses, an adaptive time-varying rms filter may be required if this trend continues, but for now we used a constant threshold, and accepted a slightly higher rejection fraction in the later parts of the mission.

Although they are still processed, data taken in staring mode are not used in the construction of the final sky mosaic images as they contribute a far higher level of systematic noise than the standard dithered observations (although this effect is less pronounced from OSA 9 onwards). Some 1290 science windows in the input dataset were flagged as consisting of staring data, representing a further  exposure loss of 3.2 Ms (2.5\%) in map construction.

After removal of high-rms and staring data, approximately 67000 scws remained in the dataset, with a total exposure of $\sim$110 Ms. The selected science windows were then combined using a proprietary image mosaic tool which statistically averages the images from multiple input maps. This process has been optimised to allow the creation of all-sky maps based on large numbers of input science windows. Mosaics were constructed for five energy bands (see Section \ref{pipeline}) with 2.4$'$ pixel resolution, significantly oversampling the intrinsic system PSF. Mosaics were made in four projections: centred on the Galactic Center,  on the Galactic anti-center, north Galactic polar and south Galactic polar. These multiple projections are intended to present the automatic source detection algorithms with source PSFs with the minimum possible distortion.

Previous catalogs have employed various timescales on which mosaics were constructed in an attempt to optimise the detection of new sources with a variety of duty cycles. We have simplified our approach, and initially constructed mosaics on only revolution  and whole-archive timescales. Revolution maps are optimised to detect sources active on timescales of the order of a day and persistent sources can best be detected in an {\em all-archive} accumulation of all available high-quality data. 

\subsubsection{Candidate list construction}
 
Maps were searched with two different algorithms, one the standard {\em SExtractor} tool \citep{sextractor}, the other designed specifically to compensate for the varying levels of systematic background found in INTEGRAL/IBIS mosaics. In total, 60 all-sky maps (and variants) and over 19000 revolution maps were constructed and searched. An initial candidate source list was created by iteratively merging the excess lists from each map into a base list which took the cleaning catalog  as a starting point. In this way, merging commences with the best reference positions for each source, and the process also ensures that all previously declared sources are checked for their presence in the new dataset. A merge radius of 8$'$ was used, and a new candidate was added to the base list if it exceeded a detection threshold of 4.5$\sigma$ in an all-archive map, or 6$\sigma$ in a revolution map, and could not be associated with an already listed source. The higher threshold for revolution maps is essential to remove false excesses caused by noise in these lower exposure maps. In addition to this higher threshold for revolution map excesses, a number of revolutions\footnote{Revolution dates can be found at http://www.cosmos.esa.int/web/integral/schedule-information} were excluded from this process due to high noise levels associated with solar activity, these being 124--129  (inclusive), 217--218, 234, 252--254, 276--277, 315, 341--342, 349, 352--356, 506--509. This process resulted in a list of 3759 excesses which was manually inspected to ensure that blended sources flagged in previous catalogs survived the merging process.

\subsection{Final source list construction}

Light curves were constructed for every candidate source in the five standard energy bands. A search for variable source emission was then performed on those light curves by using the `bursticity' method - i.e. identifying the time window within which the source significance was optimised. Time windows in the range 0.5 days up to the full duration of the light curve were tested. Once the optimum detection time window was determined, an additional map - the `burst map' - was constructed by mosaicking only those scw falling in the best time interval and using the energy band established by 'bursticity'. This method optimises the detection of any {\em known or suspected} source that emits on any timescale longer than a science window.  Following this procedure, an improved significance has been obtained for $\sim$200 sources.  

\begin{figure}[htbp]
  \centering
   \includegraphics[width=0.99\columnwidth]{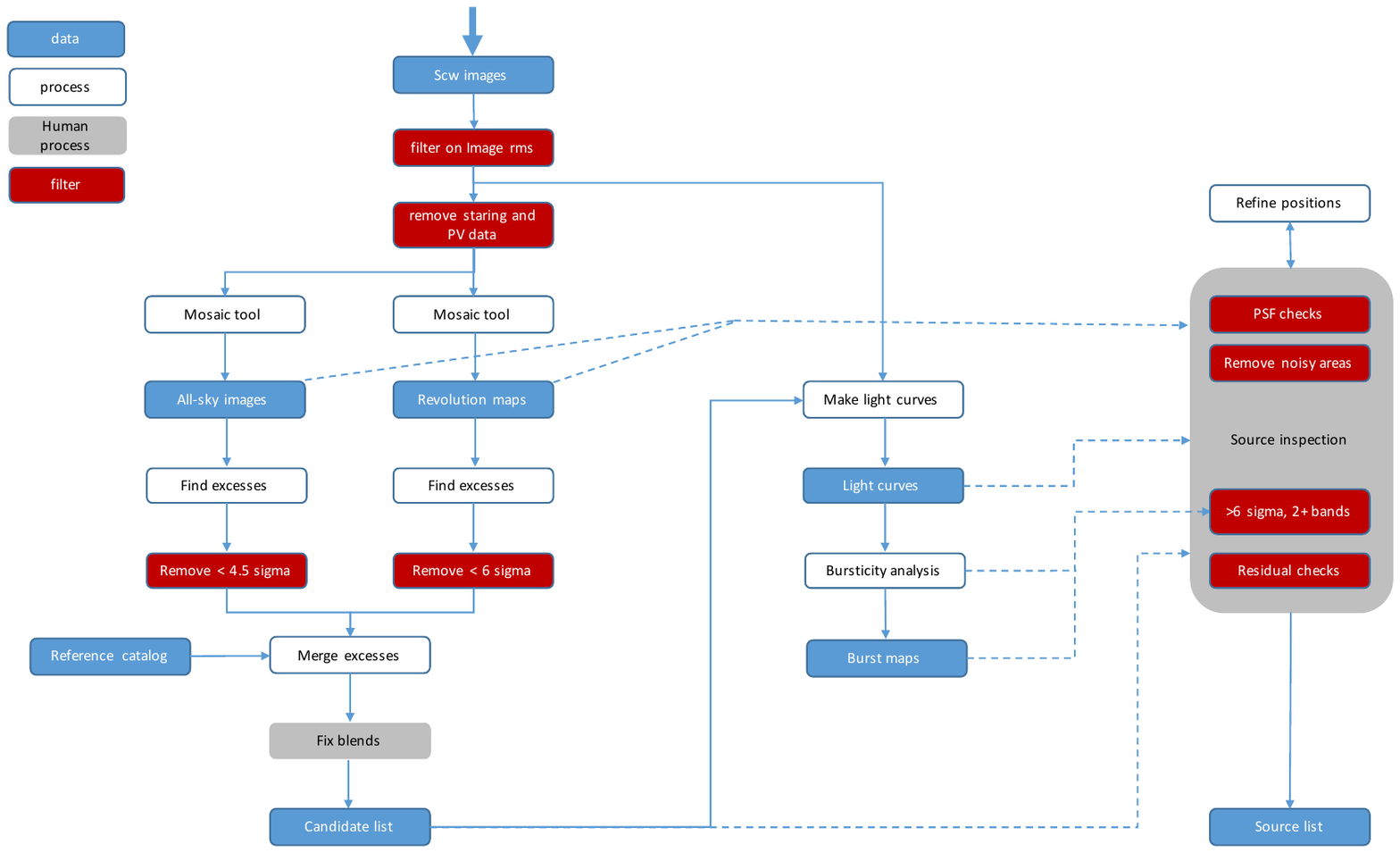}
   \caption{Data analysis and source selection flowchart, showing the filtering criteria applied at each stage. \label{fig:flow}}
\end{figure}

The final source list filtering was carried out manually. Experienced operators were presented with all the relevant data - as well as visual inspection of the maps themselves, derived parameters such as persistent significance in five energy bands, burst significance and timescales in five energy bands, local systematic levels, local image residual levels, and the total number of maps each source was detected in were quantified. A final acceptance of each putative source was made on the basis of this overall data. The overall flow of data through the analysis chain is shown in Figure~\ref{fig:flow}, which also shows the selection/rejection criteria applied at each stage.

\subsection{False positive rates}
\label{sec:FPR}

The false positive rate (FPR; the fraction of `fake' sources in the catalog) is a key parameter, and depends strongly on the methods used to identify, examine and verify the excesses. 

\begin{figure}[htbp]
  \centering
   \includegraphics[width=0.9\columnwidth]{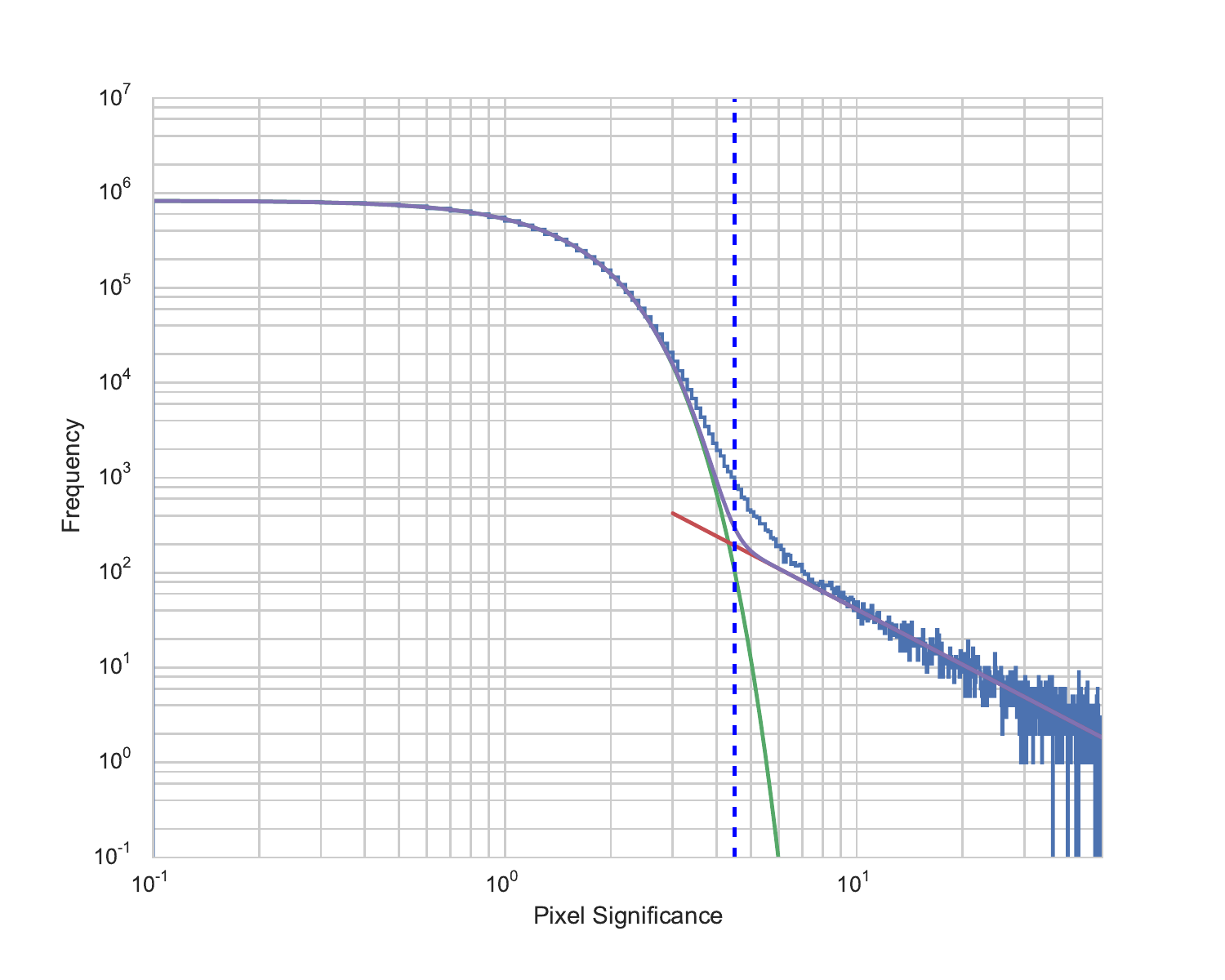}
   \caption{Pixel significance distribution for the 18-60 keV all-sky mosaic significance map. The distribution is modelled as a sum of noise (green) and source (red) contributions. The dashed vertical line is at 4.5$\sigma$. \label{fig:persistFPR}}
\end{figure}

The false positive rate for persistent sources found in inspection of IBIS mosaic images has been well established for previous catalogs \citep{Bird2004cat1,Bird2006cat2,Bird2007cat3,Bird2010cat4,Krivonos2010} and can be quantified by inspection of a histogram of either the pixel significance values in the mosaics, or the significances of the detected excesses. We follow the method of cat4 and fit the pixel distribution (Figure~\ref{fig:persistFPR}) with a Gaussian noise component and a power law component representing the sources. The point at which the noise population contributes $\sim$1\% of the source population can therefore be estimated at between 4.5 and 5 sigma, and these values have typically been used in prior catalog constructions. In this work, the threshold for 1\% FPR is 4.8 sigma, the same as that quoted for cat4, while above the formal 4.5 sigma threshold, a total of 2.6\% of the sources may be due to the noise component. We note however, that in the significance range between 4.5 and 4.8 sigma, the fraction of false sources may be as much as 25\% and have indicated this in the table with a WARN flag.

The false positive rate arising from the bursticity method is much harder to quantify. A large number of trials are performed each time a light curve is tested, and the confidence levels for any 'detection' must therefore be assessed carefully. An analytical approach to this is unlikely to yield a satisfactory result, as the $\sim$3000 light curves tested are of markedly different lengths and temporal structures (the data gaps come from the observing strategy of the telescope). Since both length and structure of the light curve affect the number of valid trials performed, they also affect the confidence limits, and we should formally assign limits on each light curve, but this is too cumbersome, and we adopt a simulation approach on the ensemble of light curves. 

We created new light curves by randomisation of existing light curves that were selected to have no source signal. For a light curve containing N data points, N swaps of (time, flux/error) pairs were performed to randomise the light curve while retaining the original overall time structure.  The advantage of this method is that a very large number of random light curves can be generated. However, the assumption in this method is that the noise in the light curve is purely statistical white-noise, as any correlated noise would be removed by the randomisation. 
 
\begin{figure}[htbp]
  \centering
   \includegraphics[width=0.8\columnwidth]{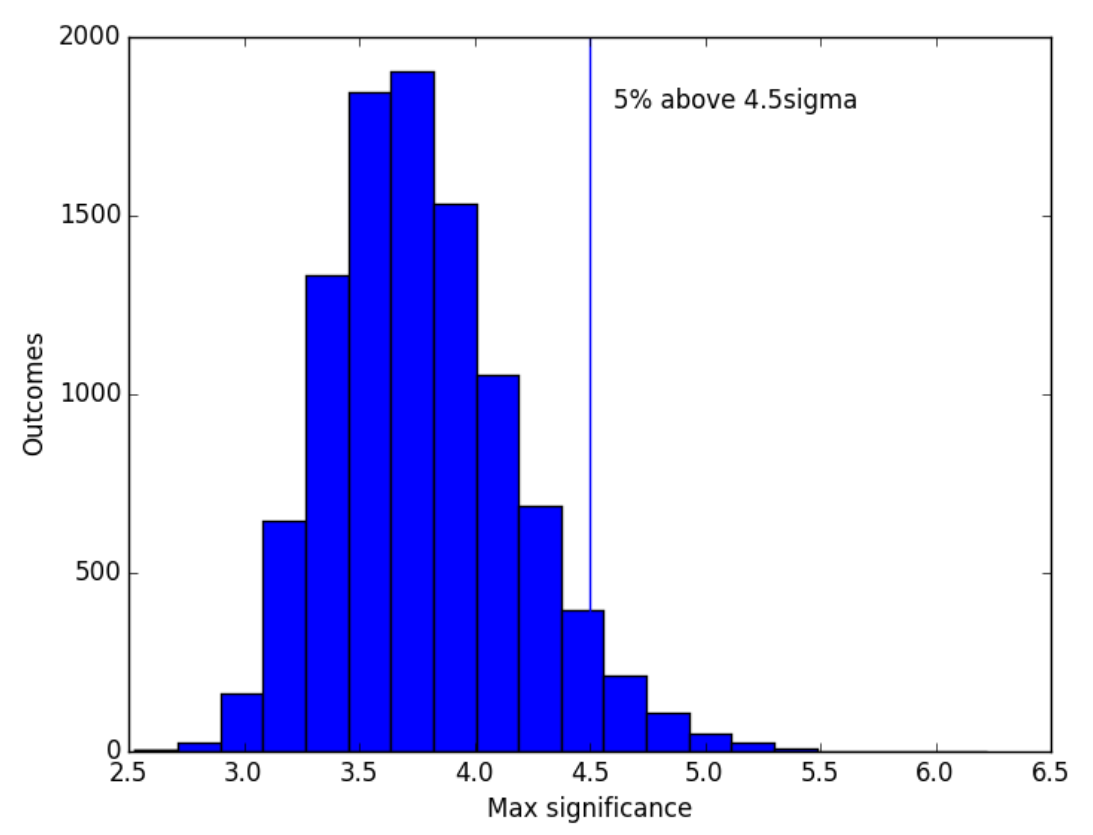}
   \caption{Maximum burst significances discovered in 10000 randomly generated light curves containing 2650 science windows with a realistic time structure and noise distribution. \label{fig:bursticityFPR}}
\end{figure}

The distribution of burst significances detected in 10000 randomised light curves derived from a medium length (2650 scw) light curve is shown in Figure~\ref{fig:bursticityFPR}.  For this typical light curve length, 10\% of the 'bursticity' tests resulted in a detection above 4.5 sigma. Corresponding values for short (550 scw) and long (10700 scw) light curves were 0.1\% and 10\% respectively.  In all these tests, less than 1\% of iterations generated a detection above 6 sigma.

We performed a second set of simulations based on inversion of the $\sim$3000 excess light curves. Each source flux light curve was subjected to a {\em sigma clipping} algorithm to remove bright positive detections above the 5$\sigma$ level, and then inverted about zero flux. In the assumption that the noise flux distribution from coded mask deconvolution is Gaussian distributed around  a mean of zero, this results in light curves with the same noise and time structure as the original light curves. These light curves will retain any systematic noise and also will maintain any longer-term noise structures (red noise) present in the originals, but should effectively contain no signal flux. Analysis of these light curves should represent a worst case scenario in terms of the derived false positive rate when compared to the purely statistical, white noise light curves made in the first approach. Using this approach, a `burst' of greater than 4.5 sigma is seen in 12.7\% of the light curve analyses, and a `burst' of greater than 6 sigma is seen in 3.5\% of the light curve analyses; this falls to 2\% at the 7 sigma level.

Combining the results of these two simulation approaches, we can estimate that the mean false alarm rate above a 6 sigma threshold is between 1\% (statistical best case) and 3.5\% (systematic worst case) for a single light curve.  Once we impose the additional requirement that the bursts are temporally aligned in more than one energy band, the statistical probability falls enormously due to the small burst duration compared to the overall light curve length. In a purely statistical sense, the probability falls to $<$1\% even above 4.5$\sigma$ as typical burst durations are $<$1\% of the light curve duration. However, we must caution that some of the systematic effects occasionally seen (e.g. poor ghost source removal)  will potentially generate noise simultaneously across all bands, so the final inspection processes are still vital to remove 'ghosts' and areas of noise in maps. We pessimistically assume a final 1\% false positive rate in the overall bursticity method when requiring simultaneous bursts in more than one band. Based on these simulations and our experience from cat4, we believe the false positive probability for sources detected on short timescales ($<$70 days) and at low significance is higher than the overall levels, and we have again indicated this in the table with a WARN flag. See Section~\ref{sec:missingcat4} and Figure~\ref{fig:badcat4} for further information.

Overall, we estimate the false positive rate in this catalog is $<$25\% for sources detected between 4.5 and 4.8 sigma in persistent maps, $<$1\% for sources detected above 4.8 sigma in persistent maps, and (pessimistically) 1\% for sources detected via the bursticity method. Thus we anticipate $\sim$30 false positives in a catalog of 939 sources, i.e. 3.5\%, with a roughly equal number coming from each detection method.

\subsection{Galactic Center Localizations\label{GC}}

\begin{figure}[htbp]
  \centering
   \includegraphics[width=0.7\columnwidth]{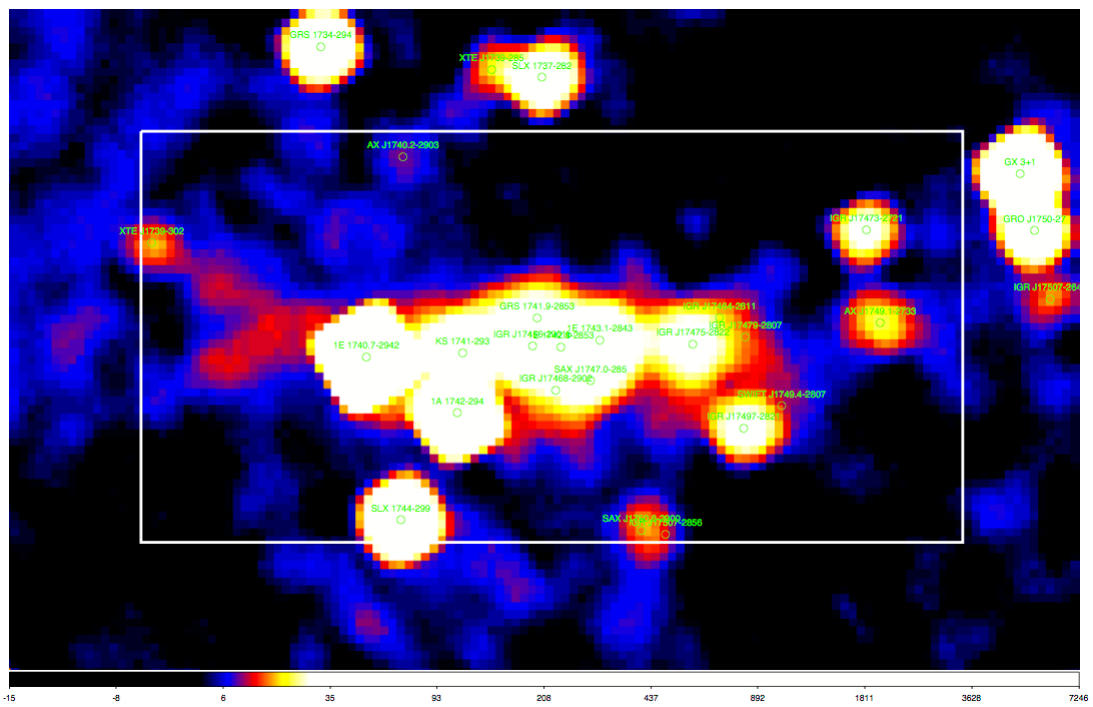}
   \caption{IBIS/ISGRI mosaic significance map of the Galactic Centre region and resulting sources from the analysis. The white box represents the central $4^{\circ} \times 2^{\circ}$ region. \label{fig:GC}}
\end{figure}

The central $4^{\circ} \times 2^{\circ}$  region of the Galaxy represents a challenging area for the INTEGRAL/IBIS map analysis. The presence of unresolved sources (and presumably many sources below the formal detection threshold creating a non-uniform background) means that the maps in this area are dominated by systematic effects and the usual statistical limits for source discovery do not apply. As a consequence, we have been extremely conservative in this region, and in fact all the sources listed are already present in the INTEGRAL Reference Catalog. Because of the complex and unresolved source distribution, the data quality for these sources may be lower than for isolated sources away from the Galactic Centre. In the source list, we indicate this with a flag (GCFLAG) with the following two values: GCFLAG=1 means that the source lies within the GC box, and is detected by our standard methods. Furthermore, the source is sufficiently resolved that we can estimate the flux and position from our maps; nevertheless we expect that the detection may be affected by nearby unresolved faint sources and the quantitative data should be treated with caution; GCFLAG=2 means that there is clear evidence of emission from the source position in one or more of our maps, but it lies within an unresolved emission region. Therefore we cannot unambiguously attribute the emission to the source, and we therefore supply the reference catalog position only. The fluxes are almost certainly contaminated by emission from nearby unresolved sources, or indeed resolved ones - in one case, two nearby sources (SAX~J1750.8$-$2900 and IGR~J17507$-$2856) are blended in stacked images of the region, but may be {\em temporally} identified as they outburst at different times, and the derived positions are unambiguously different. Nevertheless, cross-contamination of fluxes in this region is an ever-present problem. Using this approach, there are 23 sources falling within the Galactic Centre zone, of which 11 have GCFLAG=2. 

We have cross-checked our results and sources in this area with the results from the bulge monitoring project\footnote{http://integral.esac.esa.int/BULGE/} which provides a more regular monitoring and so regularly detects transient sources; nevertheless, we have no real contradictions with their database.  The differences that do exist, apart from occasional naming differences, are in fact due either to sources detected after revolution 1000 or to our detection acceptance threshold.


\section{The Table Data}
\label{sec:tabledesc}
 
The name of the source is given following the convention to quote wherever possible the name declared at the time of the first X-ray detection. The names are given in bold for the $\sim$300 sources added to the catalog since cat4. 

The astrometric coordinates of the source positions were extracted from the mosaics by the barycentring routines built into {\em SExtractor 2.5}. In almost all cases, the position for a source was extracted from the map yielding the highest source significance. In a few cases, primarily for blended sources, other maps were chosen in order to minimise the interference of other sources. Simultaneous fitting of multiple Gaussian PSFs was used in the most difficult cases - these sources are indicated as blended in the notes accompanying the table. The point source location error of IBIS is highly dependent upon the significance of the source detected \citep{Gros2003, Scaringi2010b}. We use the formulation of \cite{Gros2003}, combined with the significance of the detection used to locate the source, in order to define an error on the source position. The source localisation errors quoted are for the 90\% confidence limit.

The mean fluxes quoted in the table as $F_{20-40}$ and $F_{40-100}$ are the time-averaged fluxes over the whole dataset derived in two energy bands (20--40 and 40--100 keV). These are provided for compatibility with past catalogs and as a general reference value. However, as previously noted, their relevance as an {\em average} measure diminishes as the dataset increases and the average time of activity for many of the sources is much shorter than the on-source exposure. For variable sources, we provide a variability indicator:  a flag of Y indicates a bursticity $>1.1$ (ie a 10\% increase in significance can be obtained by selecting a single contiguous subset of the data) and a slightly variable source. A flag of YY indicates a bursticity of $>4$ (ie a 400\% increase in significance) indicating a strongly variable source. The significances quoted are the highest significance in any single map, since this gives the best indication of the robustness of source detection. However, it should be noted that the flux and significance values may derive from different energy bands and/or subsets of the data, and may initially appear contradictory. A brief commentary indicates the detection method for each source - here the term `persistent' means that the source detection is optimised in a mosaic of all data, but the detection may actually derive from a number of outbursts or flares, but {\em no single outburst} optimises the detection. For sources detected during an outburst, the MJD and duration are indicated. Warning flags are appended to some sources to indicate their position in the Galactic Center, or to warn of detections subject to higher false positive rates due to lower significance or shorter duration (see section~\ref{sec:FPR}).

The type of the source is encoded into up to four flags, which are explained in the table footnotes. We have followed the convention of \citep{Liu2007} wherever possible.  The exposure quoted is the total effective exposure on the source after all filtering of the data has been carried out.


\section{Detailed comparison with 4th IBIS/ISGRI catalog \label{sec:missingcat4}}

632 of the 723 sources detected in the 4th IBIS/ISGRI catalog are listed in this new catalog, while 4 are not included because of the new methods employed to analyse the Galactic Centre, and 87 are not included because they did not pass the new acceptance thresholds.

\begin{figure}[htbp]
\centering
\includegraphics[width=0.45\linewidth,angle=0,clip]{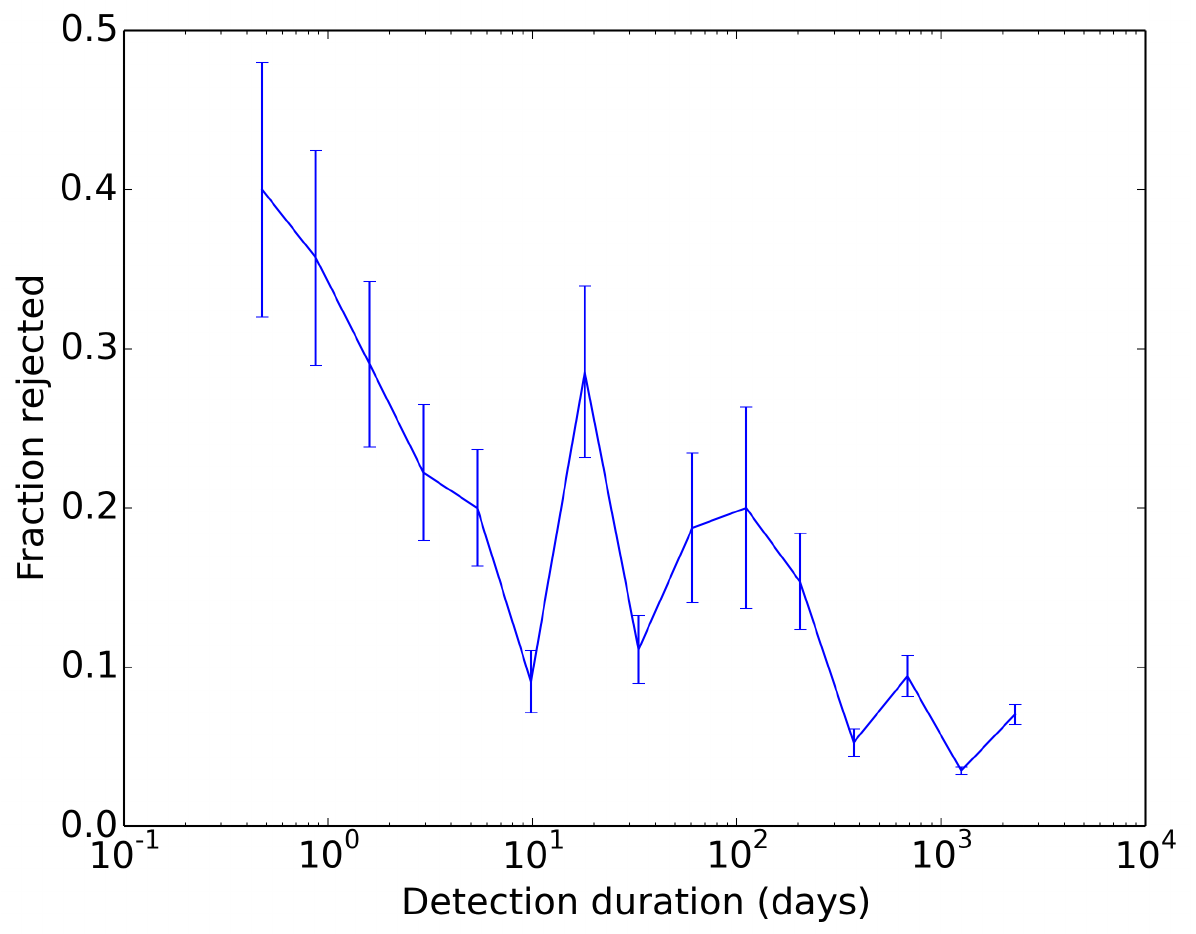}
\includegraphics[width=0.45\linewidth,angle=0,clip]{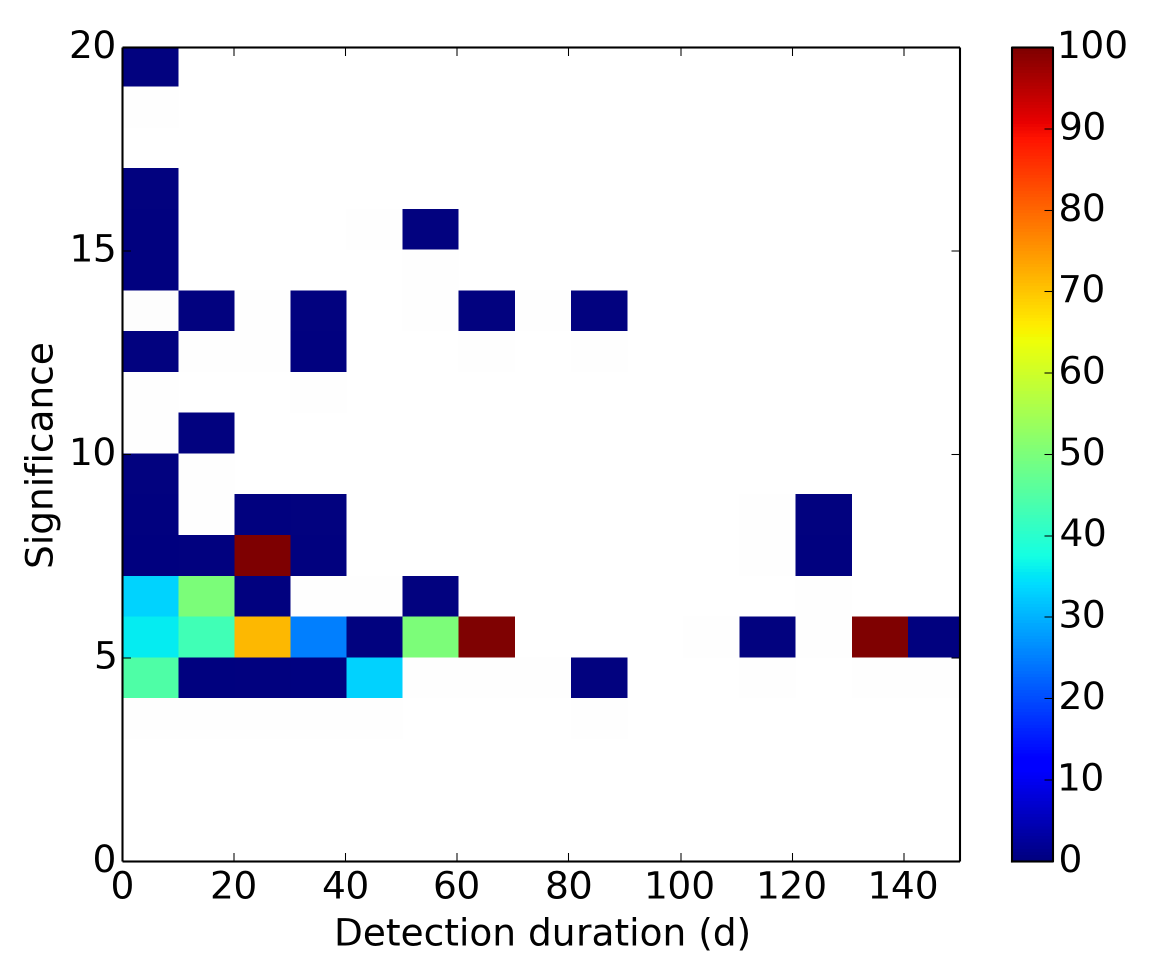}
\caption{Analysis of the false positive rates for the transient detection performed in cat4. A cat4 source is rejected as potentially a false positive if it cannot be recovered by the improved and more stringent acceptance criteria used in this work. (left) the fraction of sources rejected as a function of the source detection duration shows a clear trend towards more false positives for shorter outbursts, error bars indicate statistical uncertainties due to source numbers. (right) adding the source significance as a second parameter shows that short, low-significance outbursts contribute a very high fraction of the rejected sources.}
\label{fig:badcat4}
\end{figure}

The sources that were given in cat4, but are not detected by the cat1000 analysis have been subjected to further inspection. While we would always expect some false positives in any catalog, the number of ÔmissingÕ cat4 sources is far in excess of the expected level quoted in cat4. In total, 87 cat4 sources are not confirmed in cat1000, and breaking this subset of sources down by detection type, there is a clear trend towards these (assumed) false positives coming from the shorter outburst detections (Figure~\ref{fig:badcat4}, left). 

Further analysis of the 87 rejected sources shows that only 25 sources come from the stacked whole dataset maps, that are optimised for persistent source detection. Even then, many of these ÔpersistentÕ sources derive from low exposure (less than 200 ks) areas of the maps so may be thought of as short exposure detections. Another 14 rejected sources derive from burst maps, which may be of any exposure but tend towards shorter timescales, 31 from revolution (so $\leq 200$ ks exposure) maps, and 17 from short sequences of revolutions.  The other very clear (and expected trend) is towards low significance detections; only 12 of the missing cat4 sources were originally attributed a significance greater than 6 sigma. These trends are illustrated in Figure~\ref{fig:badcat4} (right) that shows the fraction of sources rejected between cat4 and cat1000 as a function of both duration and significance. We would expect a rejection rate of $\sim$1\% based on the expected noise content of the persistent maps, but the rejection rates for sources that were detected on short timescales ($<$70 days) and at relatively low significance ($<6\sigma$) were much higher than that; in the worst case for the short outbursts below 5$\sigma$, only 1 in 2 sources have been confirmed by the new analysis. Outside of the region bounded by duration $<$70 days and significance $<6\sigma$ the rejection fraction falls to the expected levels. These results are consistent with, and may be explained in the context of, the simulations described in Section~\ref{sec:FPR}.

For the sources that are identified in shorter periods (bursts, revolutions, sequences) we have cross-checked the outbursts detected in cat1000 against those found in cat4. In many cases there is no time correlation, and we must conclude that these cat4 excesses were probably random bright periods in the light curve of a random point of sky, and should be considered false detections. We note here that the methods employed in cat1000 are much more robust, as all five main energy bands are searched for outbursts, and we expect time correlation between the bursts in at least two of the bands. Furthermore we have operated with a much higher significance threshold for short outburst detection. However, these improved methods used in this work still only partially protect against the other likely explanation of false short bursts in cat4. Specifically, a short sequence of science windows where the data is hard to analyse due to noise, blended sources or an incomplete catalog may give rise to strong image artefacts, and we have to assume the same conditions may persist from cat4 to cat1000. Therefore detection of a burst at the same time in cat4 and cat1000, although strongly indicative, on its own is not considered 100\% confirmation of source detection. 

We must assume that those few rejected sources that were originally detected in persistent maps (i.e. by compiling all observations) and with long exposures were spurious detections of artefacts induced by the previous imaging software version, and now better suppressed in {\em OSA~9.0}. Numerous changes were implemented in the software, instrument response models and reference catalogs between {\em OSA~7.0} and {\em OSA~9.0}. The use of a newer, improved base `cleaning' catalog will certainly have played a part in reducing the image noise levels. Long-term source variability should {\em not} cause a previously known source to be rejected. Since these  previously detected sources are automatically included in our analysis, they should be detected despite a declining flux - we are confident that the bursticity analysis successfully identifies them as active during the earlier mission phases. 


Following removal of those sources not confirmed in this work, we can re-analyse the source type distribution for the confirmed cat4 sources. The modified source type distribution (Figure~\ref{fig:typedist}(left) and Table~\ref{tab:typedist}) shows the success of the various follow-up campaigns, in that only 16\% of the confirmed cat4 sources now lack association with a specific type of object. Conversely, follow-ups on the 87 ÔmissingÕ cat4 sources have largely failed to identify clear counterparts, with only 3 likely and 10 possible AGNs associations being reported. Some random correlation with sources is to be expected, and in the extragalactic sky this is most likely to produce correlations with the isotropic source populations (mostly AGNs and CVs). Nevertheless, this low rate of association with known objects gives us further confidence that we have successfully identified a subset of likely false positive excesses. 


\section{Results}
\label{sec:results}

This new catalog lists 939 sources detected with a systematic analysis of all public observations collected up to the end of 2010 and consisting of eight years of
INTEGRAL data. Of these sources, 881 have a significance above 5$\sigma$ level and can be considered more secure detections, while the rest have a lower significance in the range 4.5 to 5 $\sigma$. 

307 sources in this catalog are new entries with respect to cat4 (these are shown in bold font in the source list); some of them have been previously declared as INTEGRAL detections, or have already been included in the INTEGRAL reference catalog \citep{Ebisawa2003}.
In particular, 60 sources have previously been discovered and reported in the literature with IGR designations. A further set of 127 are already listed either in other hard X-ray catalogues, mainly that of {\em Swift}/BAT, or previously reported elsewhere. Therefore the remaining 120 sources are reported as soft gamma-ray emitters for the first time in this work.
 
\begin{figure}[htbp]
\centering
\includegraphics[width=0.9\linewidth,angle=0,clip]{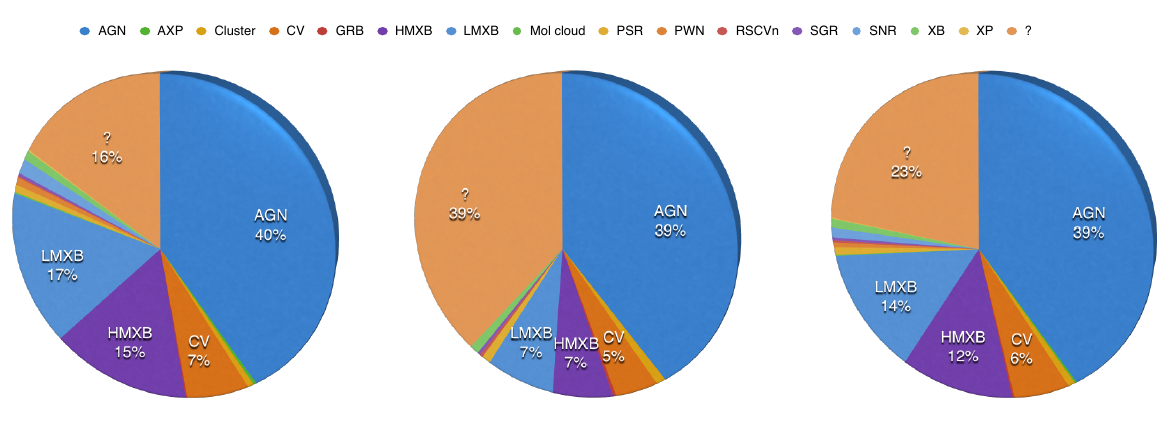}
\caption{Source type distributions for (left) confirmed cat4 sources; (center) new cat1000 sources added since cat4; (right) cat 1000 overall}
\label{fig:typedist}
\end{figure}

Table~\ref{tab:typedist} shows the source distribution by class in this new catalog - as in previous catalogs the same main source classes are detected (HMXB, LMXB, CV and AGN), including a large fraction of unassociated sources (23\%). The same data are presented graphically in Figure~\ref{fig:typedist} (center and right) 

\begin{table}[htbp]
\centering
\begin{tabular}{|l|cc|cc|cc|} \hline
Type		&	\multicolumn{2}{c|}{cat4}	& \multicolumn{2}{c|}{cat1000 new} 	& \multicolumn{2}{c|}{cat1000 overall} \\ \hline
		&	Src		& \%			& 	Src		& \%				& 	Src		& \%				\\ \hline
AGN		&	250		& 40\%		&	119		& 39\%			&	369		& 39\%			\\
?		&	100		& 16\%		&	119		& 39\%			&	219		& 23\%			\\ 
LMXB	&	106		& 17\%		&	23		& 7.5\%			&	129		& 14\%			\\
HMXB	&	96		& 15\%		&	20		& 6.5\%			&	116		& 12\%			\\
CV		&	42		& 7\%		&	14		& 5\%			&	56		& 6\%			\\
SNR		&	10		& 2\%		&	0		& $<$1\%	 		&	10		& 1\%			\\
XB		&	6		& $<$1\%		&	3		& 1\%			&	9		& 1\%			\\
PSR		&	5		& $<$1\%		&	3		& 1\%			&	8		& 1\%			\\
Cluster	&	4		& $<$1\%		&	3		& 1\%			&	7		& 1\%			\\
PWN		&	5		& $<$1\%		&	0		& $<$1\%			&	5		& 1\%			\\
SGR		&	2		& $<$1\%		&	1		& $<$1\%			&	3		& $<$1\%				\\
AXP		&	2		& $<$1\%		&	0		& $<$1\%			&	2		& $<$1\%				\\
GRB		&	1		& $<$1\%		&	1		& $<$1\%			&	2		& $<$1\%				\\
RSCVn	&	1		& $<$1\%		&	1		& $<$1\%			&	2		& $<$1\%				\\
Mol cloud	&	1		& $<$1\%		&	0		& $<$1\%			&	1		& $<$1\%				\\
XP		&	1		& $<$1\%		&	0		& $<$1\%			&	1		& $<$1\%				\\ \hline
Total		&	632		&			&	307		&				&	939          	&                       \\ \hline
\end{tabular}
\caption{Source type numbers for (left) confirmed cat4 sources; (center) new cat1000 sources added since cat4; (right) cat 1000 overall}
\label{tab:typedist}
\end{table}

Compared to cat4 (Figure~\ref{fig:typedist} (left)), there seems to be a slight change in the type fractions, i.e. overall the fraction of Galactic sources continues to reduce following the trend seen in previous catalogs. However, if we consider only the new entries (see Figure~\ref{fig:typedist}  centre panel), it is evident that there is a large fraction of sources  (39\%) that still need to be identified and among them a significant number could eventually be of a Galactic nature. 

This opens the path to a large program of follow-up work/observations  as has been successfully performed in the past. The follow-up program started ten years ago with the release of the first IBIS catalog (Bird et al. 2004) and continued thereafter. Typically, the presence of a soft X-ray source in the IBIS error box has been used to reduce the soft gamma-ray positional uncertainty and hence enable optical and NIR follow-up observations. This process has been performed either by cross-checking with a number of available X-ray catalogs (e.g. with ROSAT see \cite{Stephen2005, Stephen2006}), using IBIS itself (50\%), or using additional observations with other missions such as Swift (27\%), Chandra (17\%), XMM (5\%).

As recently reviewed by \cite{Masetti2013}, teams have so far pinpointed the nature of about 240 sources which represents a large fraction of the unidentified objects listed in all previous IBIS surveys. The majority of these sources are AGN (61\%) followed by X-ray binaries (25\%) and Cataclysmic Variables (CVs, 12\%). Most of the AGN are local Seyfert galaxies of type 1 and 2 while the largest fraction of Galactic binaries have a high mass companion. 

Overall this follow-up program has highlighted the key role played by INTEGRAL in  discovering new classes of high mass X-ray binaries (absorbed objects and supergiant fast X-ray transients), in detecting AGNs in the Zone of Avoidance i.e. the area of the sky that is obscured by
the Milky Way  \citep{Kraan2000} and at high redshifts, as well as in confirming a population of magnetic CVs emitting above 20 keV. 
Follow-up work on unassociated sources in this new catalog has already started and hopefully will lead to further identifications.

\section{Comparison with other recent soft gamma-ray catalogs}

In this section we make a brief comparison between this catalog and two other soft gamma-ray surveys. 

The first is that of \cite{Krivonos2012}, which includes a very similar dataset (up to revolution 1013) but only considers sources in the Galactic Plane. In addition \cite{Krivonos2012} used different software methods and slightly different energy bands within their data analysis to achieve a claimed identification completeness of 0.91. As well as allowing them to perform population studies on a good statistical basis \citep[see][]{Lutovinov2013}, this level of completeness provides a good comparison for this work.
 \cite{Krivonos2012}  identified 392 sources above a 5$\sigma$ detection threshold, and this number increases to 402 if the detection threshold is lowered to 4.8$\sigma$. Comparison with our catalog indicates that only 15 of their sources are not present in our list, and of those, 12 have a detection significance below 5$\sigma$ in at least one of the three energy bands used in the Krivonos catalog. The degree of agreement is thus $\sim$99\%, consistent with the statistical uncertainties associated with the two methods.
 
\begin{figure}[htbp]
\centering
\includegraphics[width=0.75\columnwidth]{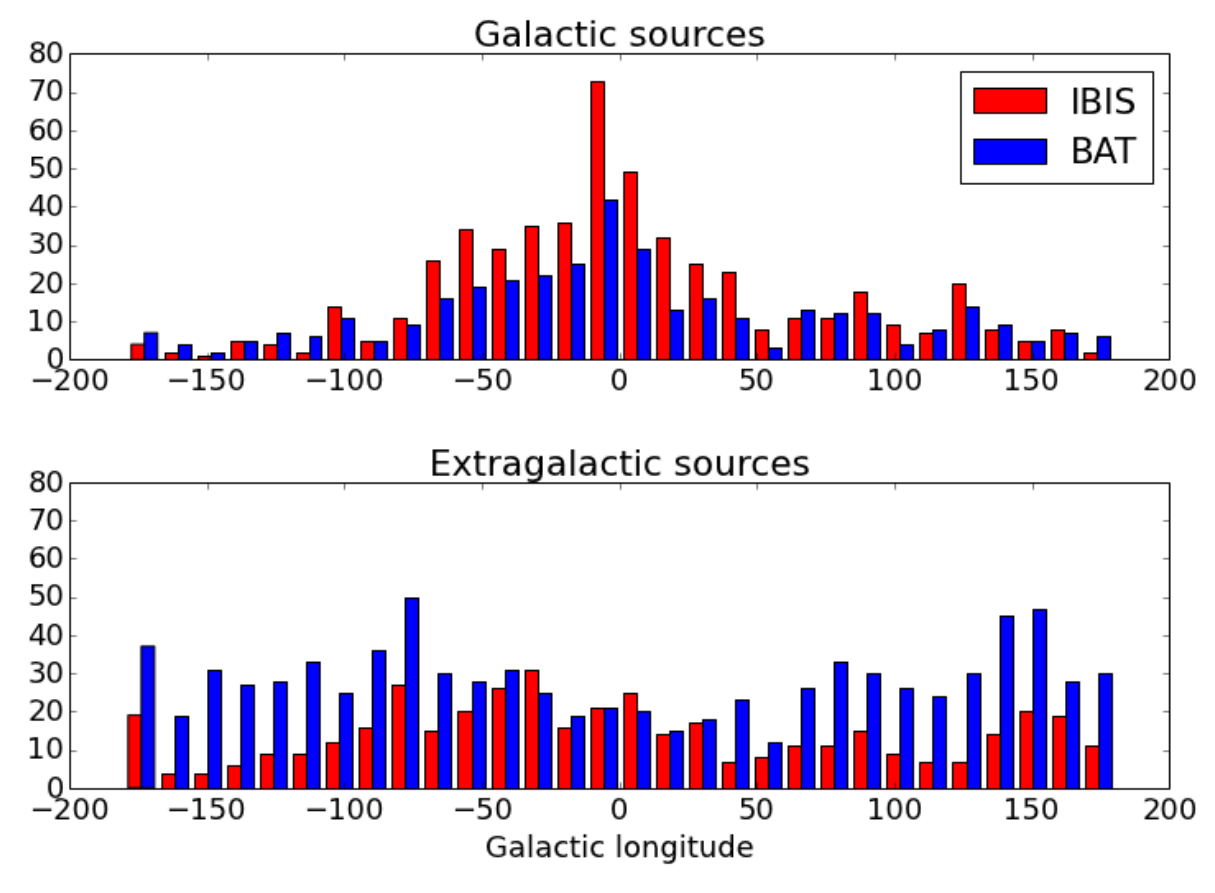}
\caption{(upper) Distribution of sources detected by INTEGRAL/IBIS (red) and {\em Swift}/BAT (blue) as a function of Galactic longitude. (upper panel) sky region within 10$^{\circ}$ of the Galactic Plane; (lower panel) sky more than 10$^{\circ}$ from the Plane.}
\label{fig:longitude}
\end{figure}

Secondly, we have also cross-correlated our source list with that of the 70 month {\it Swift}/BAT survey \citep{Baumgartner2013}. This catalog reports  1171 hard X-ray sources detected above a significance threshold of 4.8$\sigma$ in the 14--195 keV energy band. So far this is the most sensitive and uniform survey in the soft gamma-ray band, reaching a limiting flux sensitivity of $1.34 \times 10^{-11}$ erg/s/cm$^2$ over 90\% of the sky. Due to the satellite observing strategy, the {\it Swift}/BAT all-sky survey is very uniform which explains why most of the reported sources (about 60\%) are of extragalactic nature. 
Comparing the catalog of this work with that of \cite{Baumgartner2013} results in 565 correlations within a distance of 400$''$. Of these 311 are of extragalactic nature, 246 are Galactic sources and the rest lack a precise association. 
Figure~\ref{fig:longitude} shows the number of sources detected by BAT and IBIS as a function of Galactic longitude, either within 10$^{\circ}$ of the Galactic Plane (upper panel), or more than 10$^{\circ}$ from the Plane (lower panel). The differences between these two surveys is immediately apparent. It is clear that INTEGRAL has been more effective at finding soft gamma-ray emitting sources along the Galactic Plane, and particularly along the directions of the spiral arms. {\em Swift}/BAT, as a result of its larger FOV and more uniform all-sky exposure, is more effective on the extragalactic sky. Thus Figure~\ref{fig:longitude} emphasizes the great complementarity between these two missions, and the source catalogs they produce.

\subsection{Concluding comments}

As predicted in the 4th IBIS catalog, with the further data available from AO6 and AO7, this latest catalog shows a large increase in the populations in the sky beyond the Galactic Plane - seen in both the AGNs class (more than 100 new sources) and CVs. As with all previous catalogs exploiting new datasets, there is a large fraction (23\%) of unidentified sources that will require further study, and a robust follow-up programme is essential for the weak persistent examples. Transient detections represent a greater challenge for follow-up due to the serendipitous nature of their discovery. A characterisation of the transients based on their outburst duration, timing and spectral properties, and quiescent emission will be needed to further identify their nature(s). 

Once again, the soft gamma-ray sky has shown itself to be both well populated and highly variable, as the advent of missions with survey capabilities like INTEGRAL and {\em Swift} have demonstrated over the last ten years, and continue to demonstrate. Both INTEGRAL and {\em Swift} continue their highly complementary monitoring programmes, and continue to discover new sources.  Studies using long term light curves and spectral evolution are now possible and can be performed  well beyond 100 keV.  With the information reported in the current work and in combination with the recent results from the 11-years survey above 100 keV \citep{Krivonos2015} one can already derive a hardness ratio in the energy bands 20--100/100--150 keV and may understand the general properties of the 108 common sources.  We believe this will prove an invaluable data set to study the high energy behaviour of different classes of sources. By comparing these catalogue data with observations performed at lower (X-ray band) and higher (MeV and GeV bands) energies, it will be possible to create broadband spectra and therefore a unique and comprehensive view of many of the objects in our sky.


\acknowledgments

We acknowledge the founding from Italian Space agency financial and programmatic support via ASI/INAF agreement n.2013-025-R.0. ABH acknowledges support from a Marie Curie International Outgoing Fellowship within the 7th European Community Framework Programme (FP7/2007Ð2013) under grant agreement no. 275861.
 
This research has made use of:  data obtained from the High Energy Astrophysics Science Archive Research Center (HEASARC) provided by NASA Goddard Space Flight Center; the SIMBAD database operated at CDS, Strasburg, France; the NASA /IPAC Extragalactic Database (NED) operated by the Jet Propulsion Laboratory, California Institute of Technology , under contract with NASA.

\clearpage

\begin{deluxetable}{lccccccccc}
\tabletypesize{\scriptsize}
\tablecaption{IBIS Catalog of 1000 orbits \label{table:sources}}
\tablewidth{0pt}
\tablehead{ 
\colhead{Name\tablenotemark{a}} & \colhead{RA} & \colhead{Dec} &
\colhead{Error\tablenotemark{b}} & 
\colhead{F20-40\tablenotemark{c}} & 
\colhead{F40-100\tablenotemark{c}} & 
\colhead{Type\tablenotemark{d}} & 
\colhead{Vari\tablenotemark{e}} & 
\colhead{Signif\tablenotemark{f}} & 
\colhead{Exposure\tablenotemark{g}} 
}
 
\startdata
IGR J00040+7020 & 1.006 & 70.321 & 2.84 &    0.8$\pm$0.1 &    1.0$\pm$0.2 & AGN, Sy2 &  &  9.2 &   3391 \\
\hspace{.5cm}{\em \tiny  } & \multicolumn{8}{l}{\em \tiny Detected as a persistent source in the 20-100 keV band. } \\ \hline 
IGR J00234+6141 & 5.74 & 61.685 & 2.66 &    0.7$\pm$0.1 &    0.3$\pm$0.1 & CV, IP &  &  9.9 &   4777 \\
\hspace{.5cm}{\em \tiny  } & \multicolumn{8}{l}{\em \tiny Detected as a persistent source in the 17-30 keV band. } \\ \hline 
GRB041219A & 6.117 & 62.843 & 1.44 &  $<$0.1 &    0.3$\pm$0.1 & GRB & YY & 20.1 &   4812 \\
\hspace{.5cm}{\em \tiny IGR J00245+6251} & \multicolumn{8}{l}{\em \tiny Detected in a  0.6 day outburst from MJD=53358.0. } \\ \hline 
4U 0022+63 & 6.321 & 64.159 & 2.21 &    0.7$\pm$0.1 &    0.8$\pm$0.1 & SNR &  & 12.2 &   4718 \\
\hspace{.5cm}{\em \tiny  } & \multicolumn{8}{l}{\em \tiny Detected as a persistent source in the 18-60 keV band. } \\
\enddata
\tablenotetext{a}{Names in bold face indicate new detections since fourth IBIS/ISGRI catalog } 
\tablenotetext{b}{Position errors expressed as radius of 90\% confidence circle in arcminutes} 
\tablenotetext{c}{Time-averaged flux expressed in units of mCrab; appropriate conversion factors are: (20-40 keV) 10 mCrab = $7.57 \times 10^{-11}$ erg cm$^{-2}$ s$^{-1} = 1.71 \times 10^{-3}$ ph cm$^{-2}$ s$^{-1}$; (40-100 keV) 10 mCrab = $9.42 \times 10^{-11}$ erg cm$^{-2}$ s$^{-1} = 9.67 \times 10^{-4}$ ph cm$^{-1}$ s$^{-1}$} 
\tablenotetext{d}{Source type classifications: A=Atoll source (neutron star); AGN=Active galactic nuclei; AXP=Anomalous X-ray pulsar; B=Burster (neutron star); Be=B-type emission-line star; BH=Black hole (confirmed mass evaluation); BHC=Black hole candidate; BL=broad line; Cluster=Cluster of galaxies; CV=Cataclysmic variable; D=Dipping source; DN=Dwarf Nova; G=Globular Cluster X-ray source; GRB=Gamma-Ray Burst; HMXB=High-mass X-ray binary; IP=Intermediate Polar; LMXB=Low-mass X-ray binary; M=Microquasar; Mol Cloud=Molecular cloud; NL=narrow line; NS=Neutron Star; P=Polar; PSR=Radio pulsar; PWN=Pulsar wind nebula; QSO = Quasar; RG=Radio Galaxy; SFXT=Supergiant Fast X-ray Transient; SG=Supergiant; SGR=Soft gamma-ray repeater; SNR=Supernova remnant; Sy=Seyfert galaxy; Symb=Symbiotic star; T=Transient source; XB=Galactic X-ray binary; XBONG=X-ray bright, optically normal galaxy; XP=X-ray pulsar; Z=Z-type source (neutron star)} 
\tablenotetext{e}{Variability indicator, see Section \ref{sec:tabledesc} for details}
\tablenotetext{f}{Maximimum significance in a single map, see notes on individual sources for detection method.} 
\tablenotetext{g}{Corrected on-source exposure (ksec)} 

\end{deluxetable}

\end{document}